\documentclass[12pt]{article}
\usepackage{amsmath}
\usepackage{graphicx}
\usepackage{float}
\newcommand{\be}{\begin{equation}}
\newcommand{\ee}{\end{equation}}
\newcommand{\ba}{\begin{eqnarray}}
\newcommand{\ea}{\end{eqnarray}}

\newcommand{\n}[1]{\label{#1}}

\begin{document}

\title{{\bf Bayes Keeps Boltzmann Brains at Bay}
\thanks{Alberta-Thy-6-17, arXiv:YYMM.NNNNN [hep-th]}}

\author{
Don N. Page
\thanks{Internet address:
profdonpage@gmail.com}
\\
Department of Physics\\
4-183 CCIS\\
University of Alberta\\
Edmonton, Alberta T6G 2E1\\
Canada
}

\date{2017 August 1}

\maketitle
\large
\begin{abstract}
\baselineskip 18 pt

Sean Carroll has recently argued that theories predicting that observations are dominated by Boltzmann Brains should be rejected because they are cognitively unstable:  ``they cannot simultaneously be true and justifiably believed.''  While such Boltzmann Brain theories are indeed cognitively unstable, one does not need to appeal to this argumentation to reject them.  Instead, they may be ruled out by conventional Bayesian reasoning, which is sufficient to keep Boltzmann Brains at bay.

\end{abstract}

\normalsize

\baselineskip 20 pt

\newpage

\section{Introduction}

Many cosmologists now recognize that it is not sufficient just to know the dynamical laws of physics (sometimes misleadingly called the `theory of everything,' which we do not know yet, though superstring/M theory is one leading candidate that is only partially understood).  One also needs to know the quantum state (or the boundary conditions giving it), which is so far even less understood.  However, even that is insufficient, since we also need to know how to get probabilities (or perhaps more strictly, measures) of observations from the quantum state \cite{Page:2008ns}.

There are plausible postulates for some aspects of the dynamical laws and for the quantum state that, when coupled with simple ideas for the rules for getting the probabilities or measures of observations, lead to the disturbing conclusion that the measure seems to be concentrated on Boltzmann Brain observations, observations of Boltzmann Brains (BBs) that arise from relatively rapid thermal fluctuations \cite{Rees, Dyson:2002pf, Goheer:2002vf, Albrecht:2004ke, Vilenkin:2006qg, DeSimone:2008if, Carroll:2010zz} or quantum fluctuations \cite{Page:2005ur, Page:2006dt, Page:2006hr, Page:2006ys, Page:2006nt, Davenport:2010jy}, rather than observations of Ordinary Observers (OOs) that arise from relatively slow evolution.  The seemingly plausible way in which Boltzmann Brain domination arises is that Ordinary Observers are believed to require something like radiation from a nearby star (e.g., from the Sun in our case), but stars have finite lifetimes and (leaving out Boltzmann stars, which would occur much, much less frequently than Boltzmann brains) exisr only during a finite time during the history of the universe.  Therefore, Ordinary Observers can collectively exist only for a finite period of the history of the universe, whereas Boltzmann Brains can collectively exist over a much, much longer time according to the latest preferred theories about the universe, Thus the total number of Boltzmann Brains over all of spacetime can greatly dominate over the number of Ordinary Observers.  Therefore, assuming that each suitable configuration of an observer (whether a BB or an OO) leads to a comparable probability or measure for the corresponding observation, the much greater number of BBs over OOs makes the total probability or measure of BBs much greater than that of OOs.

In many simple models, the total measure of BB observations is so much greater than that of OOs that even the restricted class of BB observations that are exactly like ours has much greater measure than the OO observations exactly like ours.  Therefore, if we accept such models, we should think with high probability that we are the much more probable BBs.

Sean Carroll has recently argued \cite{Carroll:2017gkl} that although we cannot rule out such a model on statistical grounds (since it does predict that almost all observations like ours are made by Boltzmann Brains), we should rule it out for being cognitively unstable, because if we accept it and believe that our observations come from purely random thermal or quantum fluctuations, that would give us no reason for believing the model that has such random thermal or quantum fluctuations.  Therefore, as Carroll puts it, we should ``reject cosmological models that would be dominated by Boltzmann Brains (or at least Boltzmann Observers among those who have data just like ours), not because we have empirical evidence against them, but because they are cognitively unstable and therefore self-undermining and unworthy of serious consideration.''

However, for a Bayesian who believes that a Bayesian analysis should in principle be sufficient in science (assuming that one can calculate the probability of one's observation for all theories for which one assigns a significant prior probability, which of course is no mean task, and leaving aside the inevitable subjectivity of the prior probabilities), it seems a bit disturbing to have to invoke a different principle (cognitive instability) to rule out Boltzmann Brain theories.

Here I shall show that a Bayesian analysis by itself is in principle sufficient to give very low posterior probabilities to theories that predict domination by Boltzmann Brain observations.

\section{A Bayesian analysis of Ordinary Observer vs.\ Boltzmann Brain theories}

In a Bayesian analysis, if there are a set of possible observations $O_j$ and
theories $T_i$ with prior probabilities $P(T_i)$, such that the
probability of observation $O_j$ in theory $T_i$ is $P(O_j|T_i)$, then the posterior probability of theory $T_i$, given observation $O_j$, is
\be
P(T_i|O_j) = \frac{P(O_j|T_i)P(T_i)}{\sum_k P(O_j|T_k)P(T_k)}.
\n{Bayes}
\ee

Now let us suppose there are two sets of possible observations, a number $N_O$ of observations $O_j$ with $1 \leq j \leq N_O$ that are relatively ordered observations that Ordinary Observers can have, and a number $N_D$ of observations $O_j$ with $N_O+1 \leq j \leq N_O+N_D$ that are relatively disordered observations that only Boltzmann Brains have.  Since there can be thermal or quantum fluctuations to arbitrary Boltzmann Brain configurations (including configurations the same as what Ordinary Observers have), I shall assume that Boltzmann Brains can also have any of the relatively ordered observations as well.  Since presumably there are far more relatively disordered observations than ordered ones, I shall assume that $N_D \gg N_O$.  I shall further assume that we use one of our observations to test the different theories, and that this observation is a relatively ordered one, say for concreteness the observation $O_1$.

Next, suppose we have two theories to test against our observation $O_1$, theories $T_1$ and $T_2$.  Let theory $T_1$ be the theory in which there are $N_{OO}$ Ordinary Observers and no Boltzmann Brains.  Assume for inessential simplicity that each Ordinary Observer has the same number $m$ of observations, and that each of these $m$ observations for one single Ordinary Observer is distinct.  (For a human who lives about two billion seconds, this might be of the order of $10^{10}$ if there are roughly five observations per second, but the actual number does not matter.)  Assume that the universe of theory $T_1$ is sufficiently large that the number of observational occurrences, $mN_{OO}$, is the positive integer $n$ multiplied by the set of $N_O$ ordered observations, so that $mN_{OO} = nN_O$.  (The integer $n$ is the number of occurrences of each relatively ordered observation $O_j$ with $1 \leq j \leq N_O$.)  Further assume, again for inessential simplicity, that each of the $N_O$ relatively ordinary observations occurs the same number of times $n$.  

Now suppose that theory $T_1$ specifies that each occurrence of one of the $m$ observations of each one of the $N_{OO}$ Ordinary Observers makes a unit contribution of the measure to the corresponding relatively ordered observation $O_j$ with $1 \leq j \leq N_O$.  Then the total measure of all observations in this theory is $mN_{OO} = nN_O$, and each of the $N_O$ relatively ordered observations that have positive measure in this theory has measure $n$.

For Bayes' theorem, we take the conditional probability $P(O_j|T_i)$ to be the normalized measure that the theory $T_i$ assigns to the observation $O_j$, so that for each theory, the sum of the conditional probabilities or normalized measures over all observations is unity.  Therefore, for theory $T_1$, the conditional probability or normalized measure for each relatively ordered observation is $P(O_j|T_i) = 1/N_O$ (for $1 \leq j \leq N_O$), and for each relatively disordered observation is $P(O_j|T_i) = 0$ (for $N_O+1 \leq j \leq N_O+N_D$).  In particular, since theory $T_1$ is being tested by the relatively ordered observation $O_1$, its likelihood is 
\be
P(O_1|T_1) = \frac{1}{N_O}.
\n{T1}
\ee

Let theory $T_2$ be the theory that not only has $N_{OO}$ Ordinary Observers that each have $m$ observations, with these $mN_{OO} = nN_O$ total number of instances of observations equally spread among the $N_O$ relatively ordered observations so that they give a unnormalized measure $n$ for each relatively ordered observation $O_j$ with $1 \leq j \leq N_O$, but that theory $T_2$ also has $N_{BB}$ Boltzmann Brains.  Since Boltzmann Brains are short-lived thermal or quantum fluctuations, assume for inessential simplicity that each Boltzmann Brain makes only one observation, but also assume that the number $N_{BB}$ of Boltzmann Brains is some positive integer $p$ multiplied by the total number of possible observations, $N_O+N_D$, and that this number is much larger than the number of occurrences $mN_{OO} = nN_O$ of relatively ordered observations by Ordinary Observations, so that $N_{BB} = p(N_O+N_D) \gg mN_{OO} = nN_O$.  Assume that theory $T_2$ has the Boltzmann Brains contribute a total unnormalized measure $N_{BB} = p(N_O+N_D)$ that for simplicity is equally spread among all $N_O+N_D$ possible observations, so that each possible observation gains an unnormalized measure $p$ from Boltzmann Brains.

Then when we add the measure contributed by Ordinary Observers, $n$ for each of the $N_O$ relatively ordered observations, to the measure contributed by Boltzmann Brains, $p$ for each of the $N_O+N_D$ total number of observations, we get that the unnormalized measure is $n+p$ for each of the $N_O$ relatively ordered observations and $p$ for each of the $N_D$ relatively disordered observations.  The total unnormalized measure in theory $T_2$ is thus $(n+p)N_O + p N_D$.  Therefore, the normalized measure for each of the $N_O$ relatively ordered observations (including the observation $O_1$ being used to test the thoery) is $(n+p)/[(n+p)N_O + p N_D]$, and the normalized measure for each of the $N_D$ relatively disordered observations is $p/[(n+p)N_O + p N_D]$.  This implies that when tested by the relatively ordered observation $O_1$, theory $T_2$ has a likelihood
\be
P(O_1|T_2) = \frac{n+p}{(n+p)N_O + p N_D}.
\n{T2}
\ee

Eqs.\ (\ref{T1}) and (\ref{T2}) give the ratios of these likelihoods as
\be
\frac{P(O_1|T_1)}{P(O_1|T_2)} = 1 + \frac{pN_D}{(n+p)N_O}
=\left(1+\frac{N_D}{N_O}\right)
\frac{1+mN_{OO}/N_{BB}}{1+(1+N_D/N_O)mN_{OO}/N_{BB}}.
\n{likerat}
\ee
Furthermore, using Bayes' theorem Eq.\ (\ref{Bayes}), the ratio of the posterior probabilities of the Ordinary Observer theory $T_1$ and of the Boltzmann Brain theory $T_2$ is
\be
\frac{P(T_1|O_1)}{P(T_2|O_1)} = \frac{P(T_1)}{P(T_2)}
\left(1+\frac{N_D}{N_O}\right)
\frac{1+mN_{OO}/N_{BB}}{1+(1+N_D/N_O)mN_{OO}/N_{BB}}.
\n{postrat}
\ee

Surely the number $N_D$ of possible relatively disordered observations is much greater than the number $N_O$ of possible relatively ordered observations, or $N_D/N_O \gg 1$.  But initially plausible Boltzmann Brain theories tend to have the number $N_{BB}$ of Boltzmann Brains {\em enormously} larger than the number $N_{OO}$ of Ordinary Observers, and even enormously larger than the number of Ordinary Observers multiplied by the number $m$ of observations they each make during their lifetime (this $m$ being in comparison with 1 observation assumed here for each Boltzmann Brain during its typically very short lifetime), so that $N_{BB}/(mN_{OO}) \gg\gg 1$.  Therefore, I shall assume that
\be
\frac{N_{BB}}{mN_{OO}} \gg \frac{N_D}{N_O} \gg 1.
\n{ineq}
\ee
As a consequence, we get that the likelihood of the Ordinary Observer theory $T_1$ is much greater than the likelihood of the Boltzmann Brain theory $T_2$,
\be
P(O_1|T_1) \gg P(O_1|T_2)
\n{likecom}
\ee

The idea is that because Boltzmann Brains add extra unnormalized measure to the observations in a way that is less for the relatively ordered observation $O_1$ being used to test the theory than for elements of a set of other observations (the relatively disordered ones), they dilute the normalized measure of the observation $O_1$ and therefore reduce the likelihood of the theory that includes such Boltzmann Brains.

Of course, this is not the whole story, because to get the ratio of the posterior probabilities $P(T_1|O_1)$ and $P(T_2|O_1)$ as given by Eq.\ (\ref{postrat}), one must also include the ratio of prior probabilities $P(T_1)$ and $P(T_2)$.  The main reason that Boltzmann Brains have been seen as a problem is that they are predicted by what naively seems to be simpler theories than Ordinary Observer theories that avoid Boltzmann Brains, so one often takes $P(T_1) < P(T_2)$.  However, because the ratio of likelihoods is usually so much in favor of the Ordinary Observer theory, even if $P(T_1) < P(T_2)$, one ends up with the Ordinary Observer theory having the greater posterior probability:  $P(T_1|O_1) > P(T_2|O_1)$ if
\be
\frac{P(T_2)}{P(T_1)} < \left(1+\frac{N_D}{N_O}\right)
\frac{1+mN_{OO}/N_{BB}}{1+(1+N_D/N_O)mN_{OO}/N_{BB}}.
\n{ineq2}
\ee
The right hand side of this inequality is generally so large that it should not be hard to find a theory $T_1$ that is not so complex that its prior probability would reasonably be assigned so low a value that the inequality (\ref{ineq2}) would be violated.   If indeed (\ref{ineq2}) holds, then without invoking cognitive instability or needing to use it for setting the prior probabilities (which can be set by using Occam's razor to assign higher prior probabilities to simpler theories, so long as one does not make the ratio of priors on the left side of the inequality (\ref{ineq2}) larger than the right hand side), this straightforward Bayesian analysis assigns Boltzmann Brain theories, in which Boltzmann Brains strongly dominate over Ordinary Observers, lower posterior probabilities than Ordinary Observer theories, in which Boltzmann Brains do not dominate over Ordinary Observers.

Thus Bayes keep Boltzmann Brains at bay.

\section*{Acknowlegments}

I am grateful for conversations about Boltzmann Brains with a number of colleagues, such as Raphael Bousso, Sean Carroll, and Alan Guth.  I am also grateful that, so far as predicted by partial theories that seem to have the highest posterior probabilities, I have not had conversations about colleagues with Boltzmann Brains.

This research was supported in part by the Natural Sciences and Engineering Research Council of Canada, and in part by Perimeter Institute for Theoretical Physics, which hosted me during part of this work.  Research at Perimeter Institute is supported by the Government of Canada through the Department of Innovation, Science and Economic Development and by the Province of Ontario through the Ministry of Research, Innovation and Science.  I am also grateful for the hospitality of Sergei Dubovsky and Matthew Kleban at the Center for Cosmology and Particle Physics at New York University, where this paper was completed.


\end{document}